\documentclass[letterpaper]{article}
\usepackage{amsmath}

\def\art#1{[\ref{#1}]}

\begin{document}

\title{\large{Comment on the ``Excess ellipticity of hot and cold spots in the WMAP data?" by	
	Berntsen, E. and Hansen, F.K.}}
\author{A.L. Kashin and G. Yegoryan\footnote{egegham@yerphi.am}\\
{\small A.I. Alikhanian National Laboratory, Yerevan, Armenia}}

\maketitle

\begin{center}
{\bf Abstract}
\end{center}

The recent paper by Berntsen and Hansen devoted to the analysis of ellipticity of anisotropies in CMB maps, distorts some statements of previous studies, misses relevant papers, along with superficial comparison of the results (in part of definitions, the role of noise, angular resolution, model parameters).    

\vspace{0.3in}

The recent paper \art{BH} is dealing with the ellipticity analysis in the CMB maps, also in comparison with previous studies   \art{G1}, \art{G2}, \art{A}. We very briefly mention some omissions in their study.

There are distortions already while quoting statements of previous studies. E.g. they write that ``\art{G1} interpret the claimed ellipticity as evidence for geodesics mixing in a hyperbolic universe. This is contrary to reports suggesting that the universe is flat". While \art{G1} actually reads: ``since there is no way of simulating this effect, we cannot exclude that the observed behavior of ellipticity can result from a trivial topology in the popular flat $\Lambda$-CDM model, or from a non-trivial topology." 

More important is the absence of quotations to \art{GK}, where the observed ellipticity is shown to be compatible to the presence of voids in the large scale matter distribution in flat FRW Universe. Namely, while the motivation to ellipticity studies was the hyperbolicity of geodesics, flat FRW models with matter inhomogeneities can reveal that property. 

Quoting the 'theoretical mean ellipticity' 1.648 derived in \art{A}, the authors of \art{BH} neglect the fact that, that ellipticity does depend on the resolution as well as the noise level in the maps (e.g. both the resolution and noise differ in WMAP's channels, W,V). Namely, already the range of variation of the ellipticity given in \art{A}, would have worried the authors of \art{BH} with the closeness of their empirically value to that theoretical value. We do not mention the sensitivity to the definitions of the center, axes of the anisotropies.   

The empirical impact on the ellipticity of the noise level was studied in \art{G2}. As for the modeling of the effect, in \art{G1} it is noted that if there are factors not taken into account in the standard simulations, any apparent fit of a particular property of a model cannot exclude a similar even better fit to others, until more properties are studied. Among such examples can be e.g. the discussion of torus topology in \art{A} or of voids in \art{GK}.

\end{document}